\noindent\centerline{\bf A Pathway Idea in Model Building}

\vskip.3cm\centerline{A.M. MATHAI}
\vskip.2cm\centerline{Centre for Mathematical Sciences,}
\vskip.1cm\centerline{Arunapuram P.O., Pala, Kerala-68674, India, and}
\vskip.1cm\centerline{Department of Mathematics and Statistics, McGill University,}
\vskip.1cm\centerline{Montreal, Quebec, Canada, H3A 2K6}
\vskip.2cm\centerline{and}
\vskip.2cm\centerline{H.J. HAUBOLD}
\vskip.1cm\centerline{Office for Outer Space Affairs, United Nations}
\vskip.1cm\centerline{P.O. Box 500, Vienna International Centre} 
\vskip.1cm\centerline{A - 1400 Vienna, Austria, and}
\vskip.2cm\centerline{Centre for Mathematical Sciences,}
\vskip.1cm\centerline{Arunapuram P.O., Pala, Kerala-68674, India}
\vskip.3cm\centerline{\bf [Keynote Lecture delivered by Prof. Dr. A.M. Mathai]}
\vskip.5cm\noindent{\bf Abstract}
\vskip.3cm The pathway idea is a way of going from one family of functions to another family of functions and yet another family of functions through a parameter in the model so that a switching mechanism is introduced into the model through a parameter. The advantage of the idea is that the model can cover the ideal or stable situation in a physical situation as well as cover the unstable neighborhoods or move from unstable neighborhoods to the stable situation. The basic idea is illustrated for the real scalar case here and its connections to topics in astrophysics and non-extensive statistical mechanics, namely superstatistics and Tsallis statistics, Mittag-Leffler models, hypergeometric functions and generalized special functions such as the H-function etc are pointed out. The pathway idea is available for the real and complex rectangular matrix variate cases but only the real scalar case is illustrated here.

\vskip.5cm\noindent{\bf 1.\hskip.3cm Introduction}
\vskip.3cm Mathematical techniques in the area of special functions, statistical techniques in the area of statistical distribution theory and characterizations and information theory techniques in the area of generalizations of Shannon type entropies and their axiomatic definitions and properties had been developed by the first author in the period of time from 1965 to 1980 [Mathai and Rathie (1975), Mathai and Pederzoli (1977), Mathai and Saxena (1978)]. Then starting from the 1980's applications of these techniques to astrophysical problems were explored by the present authors jointly in the areas of energy generation, solar modelling, and gravitational instability problems. A summary of the findings until 1988 is available from Mathai and Haubold (1988) and later results from Mathai (1993), Mathai and Haubold (2008), and Mathai, Saxena, and Haubold (2010). It was seen that interesting results, mathematically and statistically and with potential of physical interpretations, could be obtained by the fusion of special function theory, statistical distribution theory, and information theory. The present authors' work in basic space sciences started in the 1980's and then in 1988 a sequence of United Nations Workshops, prospectively co-organized by ESA, NASA and JAXA and hosted by individual nation states was conceptualized at the Centre for Mathematical Sciences,India (http://www.cmsintl.org/).

\vskip.2cm For new results in mathematics and mathematical statistics, applications are usually found in sooner or later interplay with natural sciences and hence the vast amount of theoretical development that has been achieved, do not all have immediate applications. One sequence of results which are useful in pathway model building in physical situations based on experimentally recorded data was demonstrated recently [Haubold et al. (2012)] . An idea was introduced in the 1970's by which one could go from one family of functions of mathematical or statistical nature to another family to yet another family, and later in 2005 [see Mathai(2005), Mathai and Haubold (2007)] the idea was extended to cover real and complex scalar mathematical or random variables, as well as real and complex rectangular matrix variables. The basic notion will be explained with the help of a specific example. Consider a general input-output type situation. It could be reactions generating or annihilating particles, diffusion and transport of particles and thereby an emerging product is what is observed. Consider particle reactions and let $N(t)$ be the number density at time $t$ and the rate of reaction denoted by ${{{\rm d}N(t)}\over{{\rm d}t}}$ [Mathai and Haubold (1988)]. If the number of particles produced is proportional to the original population size then the differential equation is ${{{\rm d}N(t)}\over{{\rm d}t}}=\lambda~N(t)$ where $\lambda$ denotes the rate of reactions [Saxena et al. (2010)]. Let the diffusion rate or destruction rate be $\mu$ then the residual rate is $c=\lambda-\mu$. If production dominates then $c>0$ and if destruction dominates then $c<0$. Then for the model

$${{{\rm d}N(t)}\over{{\rm d}t}}=-c~N(t)\Rightarrow N(t)=N_0~{\rm e}^{-ct}\eqno(1.1)
$$where $N_0$ is the initial population size. If the rate of change is proportional to a power of the population size and if decay dominates then the equation and the solution are the following:
$${{{\rm d}}\over{{\rm d}t}}N(t)=-c[N(t)]^{\alpha}\Rightarrow N(t)=-[1-c(1-\alpha)t]^{{{1}\over{1-\alpha}}}.\eqno(1.2)
$$This is a power law type of behavior. For $\alpha<1$ the function in (1.2) belongs to a particular case of a type-1 beta family of functions. Let $N(t)$ in (1.2) be denoted by $N_1(t)$. For $\alpha>1$, by writing $1-\alpha=-(\alpha-1)$ and denoting $N(t)$ by $N_2(t)$, we have
$$N_2(t)=[1+c(\alpha-1)t]^{-{{1}\over{\alpha-1}}}.\eqno(1.3)
$$Here (1.3) is a special case of a type-2 beta family of functions. When $\alpha\to 1$, denoting $N(t)$ by $N_3(t)$ in this case,
$$N_3(t)=\lim_{t\to 1_{+}}N_2(t)=\lim_{t\to 1_{-}}N_1(t)={\rm e}^{-ct}.\eqno(1.4)
$$This, in fact, is the model in (1.1). Thus, $N_1(t)$ and $N_2(t)$ for $\alpha<1$ and $\alpha>1$ respectively describe a wide range of models. If the exponential form in (1.1) is the stable form in a physical situation then $\alpha$ here can be called the {\it stability parameter} and $N_1(t)$ and $N_2(t)$ can describe the unstable neighborhoods of $N_3(t)$.

\vskip.3cm\noindent{\bf 2.\hskip.3cm Optimization of Entropy}

\vskip.3cm Models in physical situations are also constructed by optimizing entropy measures. The basic Shannon entropy in a probability scheme, for the continuous situation is
$$S(f)=-k\int_{-\infty}^{\infty}f(x)\ln f(x){\rm d}x\eqno(2.1)
$$where $f(x)$ is a statistical density and $k$ is a constant. When $k$ is present, we can assume $f(x)$ to be any non-negative integrable function, need not be a statistical density. $S$ represents a {\it measure of uncertainty} in a probability scheme. If $S(t)$ is maximized over all functional $f$ satisfying the condition $\int_{-\infty}^{\infty}f(x){\rm d}x=1$ and $f(x)\ge 0$ for all $x$ then $f$ is the uniform density. If (2.1) is maximized subject to two conditions (i): $\int_{-\infty}^{\infty}f(x){\rm d}x=1$ and (ii): $E(x)$ is a given quantity, $E(x)=\int_{-\infty}^{\infty}x~f(x){\rm d}x =$ the expected value or the mean value of $x$ then we end up with $f$ being an exponential density. With reference to the number density in (1.1)- (1.3) the second condition will imply that the expected number, $E[N(t)]$ in a unity space in unit time is a fixed quantity which can also be interpreted as the {\it principle of conservation of energy} if we are dealing with energy generation. If, further, the second moment $E(x^2)$ is also fixed then we end up with the Gaussian or normal density.

\vskip.2cm The basic entropy measure in (2.1) is generalized in different directions. A class of $\alpha$-generalized entropies, their characterizations and properties are available from the book: Mathai and Rathie (1975). One of the $\alpha$-generalized entropies, in the continuous case is
$$M_{\alpha}(f)={{[\int_{-\infty}^{\infty}(f(x))^{2-\alpha}{\rm d}x-1]}\over{\alpha-1}},\alpha\ne 1,\alpha\le 2.\eqno(3.2)
$$Consider the optimization of (3.2) subject to the conditions
$$\eqalignno{(a):&~~ \int_{-\infty}^{\infty}|x|^{\delta}f(x){\rm d}x=k_1<\infty,\cr
(b):&~~ \int_{-\infty}^{\infty}|x|^{\gamma+\delta}f(x){\rm d}x=k_2<\infty\cr}
$$where $k_1$ and $k_2$ are fixed, and the optimization is done over all non-negative integrable functions. $\gamma=0,\delta=1$ is the case leading to (1.1) to (1.3) or Tsallis statistics [Tsallis (1988)]. Consider the function $g(f)$ over all functional $f$, where
$$g(f)=[f(x)]^{2-\alpha}-\lambda_1~|x|^{\gamma}f(x)+\lambda_2~|x|^{\gamma+\delta}f(x)
$$where $\lambda_1$ and $\lambda_2$ are Lagrangian multipliers. Then the Euler equation is given by
$$\eqalignno{{{\partial}\over{\partial f}}g(f)=0&\Rightarrow (2-\alpha)[f(x)]^{1-\alpha}-\lambda~|x|^{\gamma}+\lambda_2~|x|^{\gamma+\delta}=0\cr
&\Rightarrow f(x)=c_1|x|^{\gamma}[1-a(1-\alpha)|x|^{\delta}]^{{{1}\over{1-\alpha}}}&(3.3)\cr}
$$where ${{\lambda_1}\over{2-\alpha}}$ is taken as $c_1$ and ${{\lambda_2}\over{\lambda_1}}$ is taken as $a(1-\alpha),a>0$. Note that (3.3) for $\alpha<1,a>0,\delta>0, x>0$ can be called an extended generalized type-1 beta model. For $\alpha>1$, writing $1-\alpha=-(\alpha-1)$, (3.3) reduces to the following:
$$f_2(x)=c_2|x|^{\gamma}[1+a(\alpha-1)|x|^{\delta}]^{-{{1}\over{\alpha-1}}},\alpha>1,\delta>0,a>0.\eqno(3.4)
$$ Note that (3.4) can be called an extended generalized type-2 beta model. Denoting $f(x)$ under $\alpha<1$ as $f_1(x)$ we have
$$f_3(x)=\lim_{\alpha\to 1_{-}}f_1(x)=\lim_{\alpha\to 1_{+}}f_2(x)=c_3|x|^{\gamma}{\rm e}^{-a|x|^{\delta}}\eqno(3.5)
$$which can be called an extended generalized gamma model. This is the entropic pathway. If $f_1(x)$, $f_2(x)$ of (3.3)-(3.5) are taken as statistical densities then $c_1,c_2,c_3$ can act as the normalizing constants, which are available by integrating out in (3.3),(3.4) and (3.5) respectively.

$$\eqalignno{c_1&={{[a(1-\alpha)]^{{{\gamma+1}\over{\delta}}}}\over{2}}{{\Gamma({{\gamma+1}\over{\delta}}
+{{1}\over{1-\alpha}}+1)}\over{\Gamma({{\gamma+1}\over{\delta}})\Gamma({{1}\over{1-\alpha}}+1)}},\alpha<1,a>0,\delta>0,\gamma+1>0&(3.6)\cr
c_2&={{[a(\alpha-1)]^{{\gamma+1}\over{\delta}}}\over2}{{\Gamma({{1}\over{\alpha-1}})}
\over{\Gamma({{\gamma+1}\over{\delta}})\Gamma({{1}\over{\alpha-1}}-{{\gamma+1}\over{\delta}})}},\alpha>1&(3.7)\cr
&a>0,\delta>0,\gamma+1>0,{{1}\over{\alpha-1}}-{{\gamma+1}\over{\delta}}>0\hbox{  and  }\cr
c_3&={{a^{{\gamma+1}\over{\delta}}}\over{2\Gamma({{\gamma+1}\over{\delta}})}},a>0,\delta>0,\gamma+1>0.&(3.8)\cr}
$$The model in (3.3) for a general $\alpha$ is the scalar version of the pathway model of Mathai (2005). This is the distributional pathway. Here $\alpha$ is called the pathway parameter. When $\alpha<1$ then the model describes the whole family of functions belonging to extended generalized type-1 beta family. When $\alpha>1$ then we move into the whole family of functions belonging to the extended generalized type-2 beta family. When $\alpha\to 1$ then both these families go into the family of extended generalized gamma family. We can also look into the transitions in the corresponding differential equations. This is the differential pathway. Thus we have the following pathways [(Mathai and Haubold (2007)]:
\vskip.2cm
\noindent Entropic pathways
\vskip.2cm
\noindent Distributional pathways
\vskip.2cm\noindent Differential pathways.
\vskip.2cm Note that (3.3) for $x>0,\gamma=0,\delta=1,a=1$ is Tsallis statistics of non-extensive statistical mechanics which works for all the cases of $\alpha<1,\alpha>1,\alpha\to 1$. This particular case of (3.3) is  also the model in (1.2). It is said that over 5000 articles are produced on Tsallis statistics so far since it was introduced in 1988 [Tsallis (1988), see also Hamza (2005)]. It is also said that over 3000 people are working on this model giving various types of interpretations in various fields.
\vskip.2cm Model (3.3) for $\alpha>1,\delta=1,a=1,x>0$ is what is known in the literature as superstatistics [Beck and Cohen (2003), Beck (2006)]. Note that since superstatistics assumes the functional form in (3.4) for $\alpha>1$, from superstatistics one cannot get (3.3) for $\alpha<1$. In the family of pathway models, superstatistics is derived from the case $\alpha>1$ and $\alpha\to 1$ whereas Tsallis statistics covers all cases $\alpha<1,\alpha>1,\alpha\to 1$ but the main restriction here is that $\gamma=0$ or the factor $x^{\gamma}$ is absent in Tsallis model. In superstatistics $x^{\gamma}$ is present but it covers only the type-2 beta  $(\alpha>1)$ and gamma  $(\alpha\to 1)$ families of functions and not type-1 beta $(\alpha<1)$ families of functions.
\vskip.2cm In the actual applications, $x$ could be time, energy, velocity etc and then the models can have different interpretations in different disciplines.

\vskip.3cm\noindent{\bf 4.\hskip.3cm Bayesian Procedure}

\vskip.3cm The model in (3.5) for a prefixed parameter $a$ can be written as a conditional density of the type
$$f_4(x|a)={{a^{{\gamma+1}\over{\delta}}}\over{2\Gamma({{\gamma+1}\over{\delta}})}}|x|^{\gamma}{\rm e}^{-a|x|^{\delta}},a>0,-\infty<x<\infty.\eqno(4.1)
$$Suppose that the parameter $a$ has a prior density given by
$$g(a)={{1}\over{\eta^{\epsilon}\Gamma(\epsilon)}}a^{\epsilon-1}{\rm e}^{-{{a}\over{\eta}}},a>0,\eta>0,\epsilon>0\eqno(4.2)
$$where $\epsilon$ and $\eta$ are known constants. Then the unconditional density of $x$ is given by
$$\eqalignno{\int_{a}f_4(x|a)g(a){\rm d}a
&={{|x|^{\gamma}}\over{2\eta^{\epsilon}\Gamma(\epsilon)\Gamma({{\gamma+1}\over{\delta}})}}\int_{a=0}^{\infty}
a^{{{\gamma+1}\over{\delta}}+\epsilon-1}{\rm e}^{-a({{1}\over{\eta}}+|x|^{\delta})}{\rm d}a\cr
&={{|x|^{\gamma}\Gamma({{\gamma+1}\over{\delta}}+\epsilon)}\over{2\Gamma({{\gamma+1}\over{\delta}})\eta^{\epsilon}
\Gamma(\epsilon)}}[{{1}\over{\eta}}+|x|^{\delta}]^{-({{\gamma+1}\over{\delta}}+\epsilon)}\cr
&={{|x|^{\gamma}\Gamma({{\gamma+1}\over{\delta}}+\epsilon)\eta^{{\gamma+1}\over{\delta}}}
\over{2\Gamma({{\gamma+1}\over{\delta}})\Gamma(\epsilon)}}[1+\eta |x|^{\delta}]^{-({{\gamma+1}\over{\delta}}+\epsilon)}.&(4.3)\cr}
$$This (4.3) for $x>0$ is the superstatistics. Note that for the convergence of the integral ${{1}\over{\eta}}+|x|^{\delta}$ must remain positive. Hence superstatistics can only produce type-2 beta family of functions when considering gamma type conditional density for $x|a$ and gamma type marginal density for $a$. When $\eta$ is of the form $b(\alpha-1),b>0,\alpha>1$ and ${{\gamma+1}\over{\delta}}+\epsilon={{1}\over{\alpha-1}}$ then we have the pathway model for $\alpha>1$. The unconditional density of $x$ in (4.3), denoted by $f_x(x)$, can also be interpreted the following way: $f_4(x|a)$ is the density of $x$ where $a$ is a parameter. Then we are superimposing another density $g(a)$ on the density $f_4(x|a)$ and then the resulting density $f_x(x)$ can be called superimposed statistics or superstatistics. Apparently when superstatistics was introduced they were unaware of Bayesian procedures in Probability/Statistics. In Bayesian procedure, superstatistics is the unconditional  density of $x$ when $x$ and the parameter $a$, for which a prior density is assumed, both belong to gamma family of densities. A more general family of unconditional densities is available from Mathai and Haubold (2007). Dozens of papers are published on superstatistics and it is being hotly pursued in different disciplines.

\vskip.3cm\noindent{\bf 5.\hskip.3cm Fractional Considerations}
\vskip.3cm Going back to our basic growth-decay problem where the rate of change is proportional to the population size, our basic differential equation, equation (1.1), is
$${{{\rm d}}\over{{\rm d}t}}f(t)=-c~f(t),c>0\Rightarrow f(t)-f_0=-c\int ~f(t){\rm d}t.\eqno(5.1)
$$If the total integral is replaced by a fractional integral of the Riemann-Liouville type let us see what happens. The left sided Riemann-Liouville fractional integral operator is denoted by ${_0D}_{x}^{-\alpha}={_0I}_x^{\alpha}$ and it is defined as
$${_0D}_x^{-\alpha}f={{1}\over{\Gamma(\alpha)}}\int_0^x(x-t)^{\alpha-1}f(t){\rm d}t,\Re(\alpha)>0.\eqno(5.2)
$$Fractional integral can be given many interpretations in statistical literature as fraction of a total integral, as the density of residual variable $u=x-y$ where $x$ and $y$ are independently distributed real positive random variables such that $x-y>0$ etc [Mathai (2010)]. If the total integral in (5.1) is replaced by fractional integral of (5.2) then the equation becomes
$$f(x)-f_0=-c({_0D}_x^{-\alpha}f)(x)\eqno(5.3)
$$where $f_0$ is a constant. One simple way of solving this equation is by taking Laplace transforms on both sides. Let the Laplace parameter be $s$. Let the Laplace transform of $f$ be denoted by $ \tilde{f}(s)$. Then
$$\eqalignno{L_f(s)-f_0\int_0^{\infty}{\rm e}^{-sx}{\rm d}x&=-c\int_{x=0}^{\infty}{\rm e}^{-sx}[{{1}\over{\Gamma(\alpha)}}\int_0^x(x-t)^{\alpha-1}f(t){\rm d}t]{\rm d}x.\cr
\noalign{\hbox{Then}}
\tilde{f}-{{f_0}\over{s}}&=-s^{-\alpha}\tilde{f}(x)\Rightarrow\tilde{f}={{f_0}\over{s[1+cs^{-\alpha}]}}&(5.4)\cr
&=f_0\sum_{k=0}^{\infty}({{c}\over{s^{\alpha}}})^k(-1)^k.\cr
\noalign{\hbox{Taking the inverse Laplace transform we have}}
f(x)&=f_0\sum_{k=0}^{\infty}(-1)^k{{c^kx^{\alpha k}}\over{\Gamma(1+\alpha k)}}=f_0E_{\alpha}(-cx^{\alpha})&(5.5)\cr}
$$where $E_{\alpha}(\cdot)$ is the basic Mittag-Leffler function. Generalization of the basic Mittag-Leffler function are the following:
$$\eqalignno{E_{\alpha}(x)&=\sum_{k=0}^{\infty}{{x^k}\over{\Gamma(1+\alpha k)}},\Re(\alpha)>0,~E_1(x)={\rm e}^x\cr
E_{\alpha,\beta}(x)&=\sum_{k=0}^{\infty}{{x^k}\over{\Gamma(\beta+\alpha k)}},\Re(\alpha)>0,\Re(\beta)>0\cr
E_{\alpha,\beta}^{\gamma}(x)&=\sum_{k=0}^{\infty}{{(\gamma)_k}\over{k!}}{{x^k}\over{\Gamma(\beta+\alpha k)}},\Re(\alpha)>0,\Re(\beta)>0&(5.6)\cr
\noalign{\hbox{where $(\gamma)_k$ is the Pochhammer symbol}}
(\gamma)_k&=\gamma(\gamma+1)...(\gamma+k-1), (\gamma)_0=1,\gamma\ne 0.\cr}
$$More generalized form of (5.5) is the Wright's function, which is a special case of the H-function. More on the applications of these functions may be seen from Mathai and Haubold (2008), Mathai et al. (2010).
\vskip.2cm
It is seen that when we move from a total differential equation to a fractional differential equation, Mittag-Leffler function and its generalizations, Wright function and H-function enter into the solutions. A series of recent papers  are available on the solutions of fractional reaction equations and fractional reaction-diffusion equations. The Laplace transform in (5.4) belongs to a general class of Laplace transforms, see Mathai et al. (2006) and the various references therein, and various members from this general class appear when solving some fractional differential equations. Some of the papers may be seen from Haubold et al. (2011) and Saxena et al (2010).

\vskip.2cm The effects of power transformations and exponentiation on various models can be seen from a recent paper Mathai (2012). Let us see what happens if a parameter is becoming larger and larger in a  Mittag-Leffler model of (5.6). Suppose that $\beta$ is real and it is becoming larger and larger. Then by using the asymptotic expansion of gamma functions or as a first approximation the Stirling's formula
$$\Gamma(z+a)\approx \sqrt{2\pi}z^{z+a-{1\over2}}{\rm e}^{-z}\hbox{  for  }|z|\to \infty,\hbox{ $a$ is bounded}
$$we see that
$$\eqalignno{\Gamma(\beta)E_{\delta,\beta}^{\gamma}(a(\beta x)^{\delta})&\approx \sum_{k=0}^{\infty}{{a^k(\gamma)_kx^{\delta k}}\over{k!}}{{\sqrt{2\pi}\beta^{\beta-{1\over2}}{\rm e}^{-\beta}}\over{\sqrt{2\pi}\beta^{\beta-{1\over2}+\delta k}{\rm e}^{-\beta}}}\cr
&=\sum_{k=0}^{\infty}{{a^k(\gamma)_k}\over{k!}}(({{x}\over{\beta}})^{\delta})^k=(1+a x^{\delta})^{-\gamma}.&(5.7)\cr}
$$This is the pathway model, Tsallis statistics and superstatistics for the case $\alpha>1$ for $\gamma={{1}\over{\alpha-1}},a=b(\alpha-1),b>0,\alpha>1,\delta>0$. For $\gamma=1$, (5.7) becomes a power series.

\vskip.3cm\noindent{\bf 6.\hskip.3cm A Mathematical Perspective}

\vskip.3cm Mathematically speaking the whole process of transition from one functional form to another, Tsallis statistics, superstatistics and pathway models in the scalar case can be described as getting rid off some parameters from a hypergeometric series. Take for example a ${_1F_1}$ series:
$${_1F_1}(a;b;x)=\sum_{k=0}^{\infty}{{(a)_k}\over{(b)_k}}{{x^k}\over{k!}}.\eqno(6.1)
$$If we wish to get rid off an upper or lower parameter then we do a limiting process.
$$\eqalignno{\lim_{a\to\infty}{_1F_1}(a;b;{{x}\over{a}})&= {_0F_1}(~;b;x)\cr
\lim_{b\to\infty}{_1F_1}(a;b;bx)&= {_1F_0}(a;~;x),~|x|<1\cr
\lim_{a\to\infty}{_1F_0}(a;~;{{x}\over{a}})&= {_0F_0}(~;~;x)={\rm e}^x&(6.2)\cr
\lim_{b\to\infty}{_0F_1}(~;b;bx)&= {_0F_0}(~;~;x)={\rm e}^x.&(6.3)\cr}
$$The binomial function going to the exponential function in (6.2) is the basis for the pathway idea, Tsallis statistics and superstatistics. Observe that a similar rich class of pathways are available from (6.3) where a Bessel function is going to an exponential function. All the above limiting forms are available by using the fact that
$$\lim_{a\to\infty}{{(a)_k}\over{a^k}}=1 =\lim_{a\to\infty}{{a^k}\over{(a)_k}}.\eqno(6.4)
$$All these ideas are extended to the matrix-variate cases, to real positive definite, hermitian positive definite and to rectangular matrices, see the basic paper Mathai (2005), and later papers by the author and his co-workers are also available. One such model is the following:
$$f(X)=c~|A^{1\over2}XBX'A^{1\over2}|^{\gamma}|I-a(1-\alpha)A^{1\over2}XBX'A^{1\over2}|^{{{\eta}\over{1-\alpha}}}\eqno(6.4)
$$where $X$ is a $p\times r,~r\ge p$ matrix of full rank $p$ of distinct real random or mathematical variables, $A$ is a $p\times p$ constant positive definite matrix, $B$ is a $r\times r$ constant positive definite matrix, $X'$ denotes the transpose of $X$, $A^{1\over2}$ denotes the positive definite square root of the positive definite matrix $A$,  $f(X)$ is a real-valued scalar function of $X$ and $c$ is a constant. This $c$ can act as a normalizing constant if $f(X)$ is treated as a statistical density. If the matrix $X$ is relocated at some other matrix $M$ then replace $X$ by $X-M$ in the model. The constants $\eta >0,~ a>0$ and $\alpha$ are real scalars where $\alpha$ is the pathway parameter. For $\alpha <1$ the model in (6.4) will stay in the generalized real matrix-variate type-1 beta family of functions. For $\alpha>1$ the model in (6.4) will go to the generalized real matrix-variate type-2 beta family of functions. When $\alpha\to 1$ both these type-1 beta and type-2 beta families will go to a generalized matrix-variate gamma family of functions. This can be seen by using the result
$$\lim_{\alpha\to 1}|I-a(1-\alpha)A^{1\over2}XBX'A^{1\over2}|^{{{\eta}\over{1-\alpha}}}=\exp\{-a\eta~{\rm tr}(A^{1\over2}XBX'A^{1\over2})\}
$$where ${\rm tr}(\cdot)$ denotes the trace of $(\cdot)$. It can be seen that all the real matrix-variate densities that are used in the current literature are available from the model (6.4) for various values of the pathway parameter $\alpha$. A similar rich family is there if we consider the transition from a Bessel form to the exponential form. Model, corresponding to the one in (6.4), is available when the variables are in the complex domain also. The results are parallel.

\vskip.3cm\noindent{\bf Acknowledgment}

\vskip.3cm The authors would like to thank the Department of Science and Technology, Government of India for the financial assistance for this work under project No: SR/S4/MS:287/05 and the Centre for Mathematical Sciences, India, for providing all facilities.

\vskip.3cm\noindent\centerline{\bf References}

\vskip.5cm\noindent~~Beck, C. (2006):~~ Stretched exponentials from superstatistics, {\it Physica A}, {\bf 365}, 96-101.

\vskip.2cm\noindent~~Beck,C. and Cohen, E.G.D. (2003):~~ Superstatistics, {\it Physica A}, {\bf 322}, 267-275.

\vskip.2cm\noindent~~Hamza, A.M. (2005):~~On the relevance of fractal diffusion to auroral backscatter, {\it Journal of Atmospheric and Solar-Terrestrial Physics}, {\bf 67}, 1559-1565.

\vskip.2cm\noindent~~Haubold, H.J., Mathai, A.M., and Saxena, R.K. (2011):~~Further solutions of fractional reaction-diffusion equations in terms of the H-function, {\it Journal of Computational and Applied Mathematics}, {\bf 235}, 1311-1316.

\vskip.2cm\noindent~~Haubold, H.J., Mathai, A.M., and Saxena, R.K. (2012):~~Analysis of solar neutrino data from SuperKamiokande I and II: back to the solar neutrino problem, {\bf arXiv: 1209.1520}.

\vskip.2cm\noindent~~Mathai, A.M. (1993):~~{\it A Handbook of Generalized Special Functions for Statistical and Physical Sciences}, Oxford University Press, Oxford.

\vskip.2cm\noindent~~Mathai, A.M. (2005):~~A pathway to matrix-variate gamma and normal densities, {\it Linear Algebra and Its Applications}, {\bf 396}, 317-328.

\vskip.2cm\noindent~~Mathai, A.M. (2010):~~Some properties of Mittag-Leffler functions and matrix-variate analogues: A statistical perspective, {\it Fractional Calculus \& Applied Analysis}, {\bf 13(1)}, 113-132.

\vskip.2cm\noindent~~Mathai, A.M. (2012):~~Statistical models under power transformations and exponentiation, {\it Journal of the Society for Probability and Statistics}, {\bf 13}, 1-19.

\vskip.2cm\noindent~~Mathai, A.M. and Haubold, H.J. (1988):~~{\it Modern Problems in Nuclear and Neutrino Astrophysics}, Akademie-Verlag, Berlin.

\vskip.2cm\noindent~~Mathai, A.M. and Haubold, H.J. (2007):~~Pathway model, superstatistics, Tsallis statistics and a generalized measure of entropy, {\it Physica A}, {\bf 375}, 110-122.

\vskip.2cm\noindent~~Mathai, A.M. and Haubold, H.J. (2007):~~On entropic, distributional, and differential pathways, {\it Bulletin of the Astronomical Society of India}, {\bf 35}, 669-680.

\vskip.2cm\noindent~~Mathai, A.M. and Haubold, H.J. (2008):~~{\it Special Functions for Applied Scientists}, Springer, New York.

\vskip.2cm\noindent~~Mathai, A.M. and Pederzoli, G. (1977):~~{\it Characterizations of the Normal Probability Law}, Wiley Eastern, New Delhi, and Wiley Halsted, New York.

\vskip.2cm\noindent~~Mathai, A.M. and Rathie, P.N. (1975):~~{\it Basic Concepts in Information Theory and Statistics: Axiomatic Foundations and Applications}, Wiley Eastern, New Delhi, and Wiley Halsted, New York.

\vskip.2cm\noindent~~Mathai, A.M. and Saxena, R.K. (1978):~~{\it The H-function with Applications in Statistics and Other Disciplines}, Wiley Eastern, New Delhi, and Wiley Halsted, New York.

\vskip.2cm\noindent~~Mathai, A.M., Saxena, R.K., and Haubold, H.J. (2006):~~A certain class of Laplace transforms with applications to reaction and reaction-diffusion equations, {\it Astrophysics and Space Science}, {\bf 305}, 283-288.

\vskip.2cm\noindent~~Mathai, A.M., Saxena, R.K., and Haubold, H.J. (2010):~~{\it The H-function: Theory and Applications}, Springer, New York.

\vskip.2cm\noindent~~Saxena, R.K. Mathai, A.M., and Haubold, H.J. (2010):~~Solutions of the fractional reaction equation and the fractional diffusion equation, {\it Astrophysics and Space Science Proceedings 2010}, 53-62.

\vskip.2cm\noindent~~Tsallis, C. (1988):~~Possible generalizations of Boltzmann-Gibbs statistics, {\it Journal of Statistical Physics}, {\bf 52}, 479-487.

\bye